\documentclass{article}

\usepackage{PRIMEarxiv}

\usepackage[utf8]{inputenc} 
\usepackage[T1]{fontenc}    
\usepackage{hyperref}       
\usepackage{url}            
\usepackage{booktabs}       
\usepackage{amsfonts}       
\usepackage{nicefrac}       
\usepackage{microtype}      
\usepackage{lipsum}
\usepackage{fancyhdr}       
\usepackage{graphicx}       
\graphicspath{{media/}}     
\usepackage{multirow}
\usepackage{amssymb}
\usepackage{pifont}
\newcommand{\cmark}{\ding{51}}%
\newcommand{\xmark}{\ding{55}}%
\usepackage{appendix}
\usepackage[
backend=biber,
style=numeric,
sorting=none
]{biblatex}

\title{GAN Inversion for Data Augmentation to Improve Colonoscopy Lesion Classification
}

\author{
  Mayank Golhar$^1$, Taylor L. Bobrow$^1$, Saowanee Ngamruengphong$^{1,2}$, Nicholas J. Durr$^{1*}$ \\
    $^1$Johns Hopkins University, Baltimore\\
    $^2$Johns Hopkins Hospital, Baltimore \\
    $^{*}$Corresponding Author : ndurr@jhu.edu
  }
\begin{document}

\maketitle

\begin{abstract}
A major challenge in applying deep learning to medical imaging is the paucity of annotated data. This study demonstrates that synthetic colonoscopy images generated by Generative Adversarial Network (GAN) inversion can be used as training data to improve the lesion classification performance of deep learning models. This approach inverts pairs of images with the same label to a semantically rich \& disentangled latent space and manipulates latent representations to produce new synthetic images with the same label. We perform image modality translation (style transfer) between white light and narrowband imaging (NBI). We also generate realistic-looking synthetic lesion images by interpolating between original training images to increase the variety of lesion shapes in the training dataset. We show that these approaches outperform comparative colonoscopy data augmentation techniques without the need to re-train multiple generative models. This approach also leverages information from datasets that may not have been designed for the specific colonoscopy downstream task. E.g. using a bowel prep grading dataset for a polyp classification task. Our experiments show this approach can perform multiple colonoscopy data augmentations, which improve the downstream polyp classification performance over baseline and comparison methods by up to 6\%.
\end{abstract}


\section{Introduction}
Colorectal cancer (CRC) is the second leading cause of death by cancer in the United States\cite{Siegel}. Early detection and removal of CRC and its precursors via optical colonoscopy can improve the 5-year survival rate by up to 88.5\% \cite{Wiegering}. The protective value of screening colonoscopy could be improved with deep learning-based systems that assist polyp detection and classification\cite{Nogueira}. However, the training of supervised deep learning models requires a large amount of data with high-quality annotations. In medical imaging, annotating data is especially time-consuming and expensive because of the requirement of domain expertise and the issues surrounding protected health information.

One solution is to increase the training dataset size by image augmentation. While there exist many data augmentation techniques such as geometric, photometric transforms, etc.\cite{Shorten}, these rudimentary approaches do not adequately span the expected test distribution. For performing colonoscopy relevant augmentations such as imaging modality translation, Generative Adversarial Network (GAN)-based techniques have recently become popular. In \cite{Wimmer,Mathew,Mahmood}, authors use GANs to translate colonoscopy images between different imaging domains. The disadvantage of these approaches is that they require training a new model for each translation task. These methods also do not provide a fine-tuned control over desired features such as the ability to continuously vary NBI intensity or lesion size.

In this paper, we propose a GAN inversion-based method for colonoscopy specific data augmentation. This approach can \textit{edit selective attributes} of real colonoscopy images by modifying their latent representations. We can then use the edited image for data augmentation while \textit{retaining the original label}. An important advantage of this method is that it performs various data augmentation tasks using only a single model. This approach not only performs multiple domain translations but also novel augmentations such as progressively varying one attribute e.g. lesion size by linear interpolation, which is not possible with other methods. 

Our contributions are as follows: 
\begin{enumerate}
    \item To the best of our knowledge, this is the first work using GAN inversion for data augmentation in colonoscopy.
    \item We showcase two types of augmentations - imaging modality translation between white light \& narrow band imaging, and changing lesion size by interpolation. These augmentations improve the performance of the downstream polyp classification network.
    \item Our experiments present a framework on how GAN inversion can use public colonoscopy datasets to increase the diversity \& robustness of internal datasets.
\end{enumerate}

\section{Methods}
\label{sec:Methods}

\begin{figure*}[h]
    \centering
    \includegraphics[width=0.7\columnwidth]{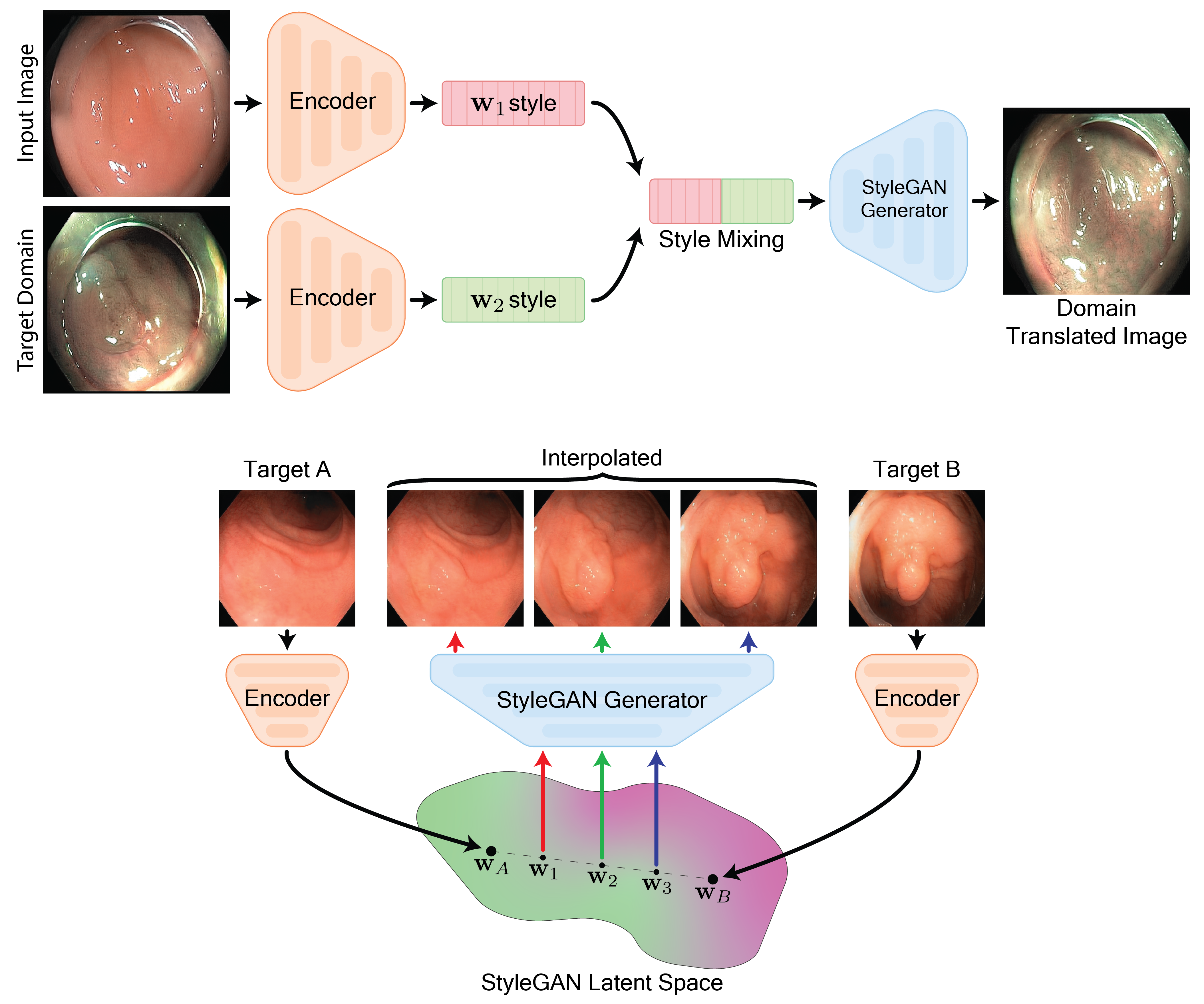}
    \caption{\textbf{Top :} Style mixing method for translating imaging domain between NBI \& WLI modes . \textbf{Bottom :} Procedure for generating images for data augmentation by interpolation in the latent space}
    \label{fig:latent_mani}
\end{figure*}

The framework of GAN inversion based data augmentation is as follows. 
Images are transformed to a latent representation by an encoder. The latent vectors are then manipulated and fed to a generator to produce an output image with different style characteristics than the original image while retaining the original image class label. Using this method, a small initial pool of labeled images can be manipulated to produce a much larger training dataset for classification tasks. We first outline the steps for training the GAN inversion model (\ref{ganinversion}), followed by the process for manipulating the latent vectors to generate new images (\ref{translating}, \ref{generating}). We then utilize the augmented dataset to train a polyp classification model (\ref{classification}).

\subsection{GAN Inversion Model}
\label{ganinversion}
The generator $(\mathcal{G})$ of a conventional GAN maps the latent representation $\mathbf{z} \in \mathcal{Z}$ to image $x$. In contrast, GAN inversion learns a mapping from an image $x$ to the latent representation $\mathbf{z}^{*} \in \mathcal{Z}$, such that when $\mathbf{z}^{*}$ is given to the trained generator $(\mathcal{G})$, it generates an inverted image $x^{*}$ that is similar to $x$ \cite{Xia}. In learning-based GAN inversion model, training can be organized into two sequential steps - training the generator, then training an encoder with the generator fixed.

For the first step, we train a StyleGAN-ADA (Adaptive discriminator augmentation) \cite{Karras_2020} model with publicly-available colonoscopy datasets.
We use the StyleGAN-ADA architecture as it produces high-quality realistic-looking images even with limited amounts of data. It can do so by using an ADA mechanism that prevents the discriminator from overfitting.
For training the model, we used public colonoscopy datasets. This was done to create a diverse training set of images captured at multiple hospitals with different colonoscopes providing a wide range of illumination conditions, GI pathologies, imaging modalities, surgical procedures, patient-specific characteristics, etc. This will enable StyleGAN \cite{Karras_2019} latent space $\mathcal{W}$ to be richly encoded with colonoscopy semantic features. 
The key idea is to train the StyleGAN with datasets whose attributes we want to transfer to the downstream training data via augmentation. Such attributes can be imaging modality, variations in pathological features such as lesion sizes, endoscope characteristics, patient-specific characteristics, etc.

For the second step, we train an encoder4editing (e4e)\cite{omer} model for GAN inversion. We chose the e4e encoder as it generates latent codes which have been shown to have high editability, \emph{i.e.}, edited images look more realistic. This is achieved by lowering the variance between the different style codes \& ensuring they lie close to the StyleGAN latent space $\mathcal{W}$. During the e4e training,  the encoder is trained while the generator weights defined in the previous step are kept fixed. We train the e4e encoder with publicly available colonoscopy datasets as well as our target dataset for the downstream task. We found that using the target dataset in the e4e training enables better reconstructions during the downstream augmentation.

\subsection{Translating Imaging Modality between WLI \& NBI} \label{translating}
In traditional GANs, the latent code $\mathbf{z} \in \mathcal{Z}$ is input to the first layer of the generator. In the StyleGAN family, the input latent code $\mathbf{z}$ is first transformed into an intermediate latent code $\mathbf{w} \in \mathcal{W}$ and then fed into different convolution layers of the generator. The intermediate latent code space $\mathcal{W}$ is shown to be less entangled \emph{i.e.} there exists linear subspaces which cause variations only in one attribute. Also, being able to input latent codes into multiple convolution layers allows for separation of features at different size scales. For \emph{E.g.} the shape of a lesion could be varied separately from its texture. This gives control over the scale-specific features during synthesis. We take advantage of both these properties for translating images between the White Light Imaging (WLI) \& Narrow Band Imaging (NBI) modalities. This is essentially an application of the ‘style mixing’ task of StyleGAN.

To translate the imaging modality of input image $x_1$ to the target domain of image $x_2$, we first generate the latent codes $\mathbf{w}_{1}$ \& $\mathbf{w}_{2}$ using the e4e encoder. To generate the translated image, we retain the $\mathbf{w}_{1}$ style codes for coarse spatial resolution layers (initial layers of the generator) and use the $\mathbf{w}_{2}$ style codes for finer spatial resolution layers (final layers of the generator) as shown in Fig. \ref{fig:latent_mani}. This is done as high-level aspects such as the shape of polyps, haustral folds, etc. are controlled by the coarse styles. Whereas imaging modality-specific aspects e.g. tissue and vasculature color, texture, etc. are low-level features controlled by fine styles.
While translating imaging modalities between the two polyp images, we ensure that both the original images have the same histological label to prevent transferring any cross-label features. The two images should also be captured with similar scope orientation, otherwise we see depth artifacts. 

\subsection{Generating Linearly Interpolated Images} \label{generating}
An advantage of using multiple public colonoscopy datasets for StyleGAN training is that the latent space has diverse \& rich semantic information. It has been shown that the StyleGAN latent space $\mathcal{W}$ is smooth and continuous, producing realistic images upon traversing in latent space\cite{Karras_2019}. We exploit these properties to generate augmented images by interpolating codes in the latent space. 
Given two target images $x_a$ \& $x_b$, we invert them to the latent space using the e4e encoder to obtain $\mathbf{w}_{a}$ \& $\mathbf{w}_{b}$. We then generate intermediate latent codes $\mathbf{w}_{i}$ by applying linear interpolation as 
\begin{equation}
\mathbf{w}_{i}=\lambda \mathbf{w}_{A}+(1-\lambda) \mathbf{w}_{B}, \lambda \in(0,1)
\end{equation}
We then pass the intermediate latent representations to the generator to obtain the interpolated images. This process is shown in figure \ref{fig:latent_mani}.
While choosing target images, we make sure that they belong to the same imaging modality, either White Light Imaging (WLI) or Narrow Band Imaging (NBI). We also ensure both the target polyp images belong to the same class. Thereby, we can assign the same histological label to the interpolated images. We assume that representations of the same class are clustered together in the latent space. The linear interpolation path is thus less likely to cross the interclass hyperplane. Even if this assumption is true for a majority of interpolated data, we will potentially observe an increase in the performance of the downstream task. Our experiment results show that this is indeed the case.

\subsection{Polyp Classification} \label{classification}
We showcase the effectiveness of the proposed GAN inversion-based data augmentation by evaluating the downstream task of lesion classification. We classify input colonoscopy lesions images into two classes - neoplastic (precancerous) or non-neoplastic. We used ResNet-18\cite{He} as the classifier with a weighted cross entropy loss. The experimental setup details are described in the next section.

\section{Experiments}
\subsection{GAN inversion}
\begin{figure*}[h]
    \centering
    \includegraphics[width=\columnwidth]{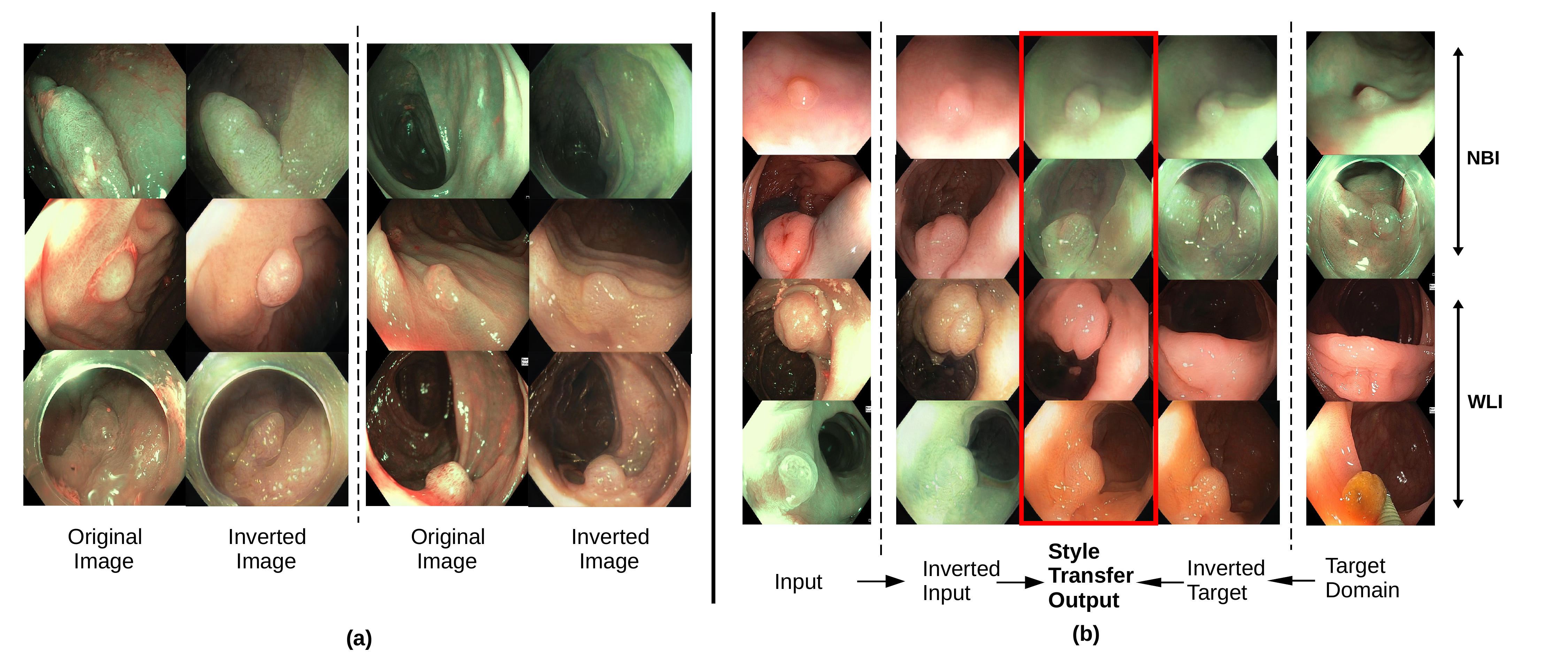}
    \caption{Qualitative results a) showing GAN inversion's reconstructions without any latent manipulations b) imaging modality translation using style mixing. }
    \label{fig:inversion_res}
\end{figure*}
For training the StyleGAN-ADA model, we use public colonoscopy datasets - HyperKvasir\cite{Borgli}, ISIT-UMR\cite{Mesejo}, and KUMC\cite{Li} datasets. In the HyperKvasir dataset, we use images of the lower GI tract \footnote{List of HyperKvasir data folders used in training is mentioned in the supplement.}. We also used all the polyp images present in the ISIT-UMR \& KUMC dataset.  A total of 72,858 images resized to 512x512 were used in training the StyleGAN. 
For the StyleGAN-ADA training, we enabled x-flips and used all augmentations except cutout. 
For all other settings, we use the default parameters. The model with the least Frechet inception distance \cite{fid} (FID = 14.16) was chosen. The total training time for the PyTorch StyeGAN-ADA model was 4 days 10 hrs on two NVIDIA RTX 3090 GPUs. Images generated from random seeds are given in the supplementary material.

In the second step, we train the e4e encoder for GAN inversion. We use the generator from the styleGAN trained in the previous step. For the training, we combine all the public datasets with the JHU dataset used for the downstream polyp classification task summing up to 74,830 images. We used the  L2 loss lambda value of 1.5 with a batch size of 6 for training models for 20000 steps. For all other parameters, we used the default values. 

 Illustrative examples showing inverted images are presented in figure \ref{fig:inversion_res} a). The results show the proposed GAN inversion model can faithfully reconstruct back colonoscopy images. As we expect, the reconstructions are not exact replicas of the input image. This is due to distortion-editability tradeoff \cite{omer} which states, we can obtain higher quality edited images by undertaking a drop in the reconstruction accuracy of the inverted image. 

\subsection{Polyp Classification Experimental Setup}
The colonoscopy dataset used for polyp classification was collected at Johns Hopkins Hospital using a protocol approved by the Johns Hopkins Institutional Review Board (\#IRB00184221). Colonoscopy procedures of 108 patients were recorded for the dataset resulting in a collection of 132 unique polyp videos. The recorded videos encompassed a wide range of imaging conditions such as imaging modalities, multiple endoscopes etc. Frames that captured polyp from multiple viewpoints were retained. A total of 1972 images were collected. The histological labels of adenoma \& serrated adenoma were assigned to the ‘neoplastic/pre-cancerous’ class whereas hyperplastic polyps were labeled as ‘non-neoplastic’. 

We evaluate our performance with a 5 -fold cross-validation scheme. For each fold, we split the dataset into train, validation, and test set in the ratio 60:20:20 at video level to prevent data leakage. Model tuning \& selection was done on validation set \& mean test metrics were reported. Class balance was maintained across sets. Training details are mentioned in the supplement.

\section{Results \& Discussion}
\subsection{Augmentation using NBI-WLI Domain Transfer}

\begin{table}[]
\caption{Results for the proposed data augmentation scheme by domain transfer between WLI and NBI modalities.}
\label{tab:translation_results}
\centering
\resizebox{0.9\columnwidth}{!}{%
\begin{tabular}{lccccccc}
\hline
Type & \textbf{Train Set} & \textbf{Test Set} & \textbf{Accuracy} & \textbf{F1-Score} & \textbf{Sensitivity} & \textbf{Specificity} & \textbf{Precision} \\ \hline
Baseline & NBI & \multirow{4}{*}{NBI} & 0.801 & 0.884 & 0.895 & 0.430 & 0.877 \\
Baseline & NBI + WLI &  & 0.787 & 0.870 & 0.865 & 0.358 & 0.878 \\
CycleGAN & NBI+$NBI_{transl}$ &  & 0.814 & 0.887 & \textbf{0.898} & 0.415 & 0.880 \\
\textbf{Ours} & NBI+$NBI_{transl}$ &  & \textbf{0.847} & \textbf{0.904} & 0.894 & \textbf{0.617} & \textbf{0.918} \\ \hline
Baseline & NBI + WLI & \multirow{3}{*}{NBI + WLI} & 0.795 & 0.878 & 0.889 & 0.285 & 0.867 \\
CycleGAN & \begin{tabular}[c]{@{}c@{}}NBI + WLI \\ + $NBI_{transl} + WLI_{transl}$\end{tabular} &  & 0.804 & 0.883 & 0.893 & 0.324 & 0.876 \\
\textbf{Ours} & \begin{tabular}[c]{@{}c@{}}NBI + WLI \\ + $NBI_{transl} + WLI_{transl}$\end{tabular} &  & \textbf{0.824} & \textbf{0.913} & \textbf{0.893} & \textbf{0.448} & \textbf{0.898} \\ \hline
\end{tabular}
}%
\end{table}

We performed two sets of experiments for testing the domain translation-based data augmentation as shown in Table \ref{tab:translation_results}. In the first set, we translated WLI images to NBI images denoted  by $NBI_{trans}$. In the second set, we translated to both the illumination modes i.e. both WLI ($WLI_{trans}$) \& NBI ($NBI_{trans}$). 

For the WLI to NBI experiments, we performed two ablation experiments with varying training sets, and one comparison with CycleGAN\cite{Zhu}. In our approach, for each image in the training set, we generated five style transferred NBI images. More than one translation was done as there exists a wide variety of imaging conditions due to manufacturer-specific NBI parameters, tissue responses, illumination, etc. helping us create a more diverse training set. We observe that augmenting the training dataset with translated NBI images improves the polyp classification over the two baselines in almost all metrics. We observe an improvement in accuracy by 6\% over the whole training set and 4.6\% with NBI only set. This is in line with the clinical practice where NBI illumination is used to improve polyp detection and classification. Our method also outperforms CycleGAN by 3.3\% in accuracy. A potential reason is CycleGAN only does one-to-one style transfer for each image, whereas our method doesn't have such restriction.

For experiments translating to both domains, we conduct an ablation study with the whole original dataset. For the augmentation, we domain transfer each image in the dataset to five WLI \& five NBI images. In this case, we again observe that data augmentation by imaging modality translations outperforms baseline methods on all metrics with a 2.9\% accuracy improvement. Our method also performs better than CycleGAN by 2\% in terms of accuracy. This shows that style transferring to multiple target images increases the heterogeneity of the training data. Overall, these results show that domain translation of imaging modalities by the proposed method not only produces realistic-looking translated images (shown in fig. \ref{fig:inversion_res}) but also helps in improving the downstream polyp classification task.

\subsection{Augmentation by Interpolated Images}

\begin{table}[]
\caption{Quantitative results comparing training with data augmentation by generating interpolated images against baselines.}
\label{tab:interp_results}
\centering
\resizebox{0.8\columnwidth}{!}{%
\begin{tabular}{lccccccc}
\hline
\textbf{Type} & \textbf{Train Set} & \textbf{Test Set} & \textbf{Accuracy} & \textbf{F1-Score} & \textbf{Sensitivity} & \textbf{Specificity} & \textbf{Precision} \\ \hline
Baseline & WLI & \multirow{2}{*}{WLI} & 0.775 & 0.856 & \textbf{0.867} & 0.233 & 0.852 \\
\textbf{Ours} & WLI + $WLI_{interp}$ &  & \textbf{0.786} & \textbf{0.868} & 0.860 & \textbf{0.415} & \textbf{0.881} \\ \hline
Baseline & NBI & \multirow{2}{*}{NBI} & 0.801 & 0.884 & 0.895 & 0.430 & 0.877 \\
\textbf{Ours} & NBI + $NBI_{interp}$ &  & \textbf{0.829} & \textbf{0.896} & \textbf{0.906} & \textbf{0.467} & \textbf{0.889} \\ \hline
\end{tabular}
}%
\end{table}

We performed two sets of experiments one for each domain - NBI, and WLI. In both the experiments, for each training image we randomly choose three other images of the same imaging modality \& class. Then for each pair of target images, we generate three interpolated images. Overall, for each training image, we generate nine interpolated images for augmentation.
For both modalities, we observe an improvement in performance over the ablation experiments as shown in Table \ref{tab:interp_results}. We observe an accuracy improvement of 2.8\% in NBI and 1.1\% in WLI cases. One reason could be interpolation exposes the classifier to images of progressive attributes such as varying lesion sizes. We hypothesize that by providing interpolated images in the training set, the classifier can construct a smoother \&  more continuous manifold due to the larger density of samples. Hence, we obtain smoother decision boundaries and more robustness to noise\cite{Pau}.
We also see a greater boost in performance when interpolating in the NBI domain as compared to the WLI domain. A potential reason is the number of NBI images is less than the number of WLI images. Hence, augmenting with interpolated images in a smaller training set leads to a larger improvement.  

\section{Conclusion}
In this paper, we proposed a method for performing colonoscopy data augmentations using GAN inversion. With GAN inversion, we can edit desired attributes of real colonoscopy images to generate augmented images. We showcased two different types of colonoscopy relevant data augmentations: 1) domain translation between white light \& narrow band imaging modes and 2) interpolated images with progressive attributes like changing the size of lesion etc. Using these data augmentations we observe a performance boost in the downstream polyp classification as shown in table. Because of the disentangled nature of the StyleGAN latent space, we can selectively manipulate the patient \& imaging specific characteristics, while preserving the class-specific semantics. This leads to increased diversity of dataset while simultaneously retaining the label of the original image. Another advantage of the proposed GAN inversion method over other generative models is that we can perform a variety of data augmentations using only one model compared to requiring separate models for each augmentation method. Using the GAN inversion framework, colonoscopy vision researchers can instill diverse attributes into their internal datasets, by leveraging information from the richly encoded StyleGAN latent space trained with public colonoscopy datasets. The next steps include exploring other GAN inversion models and applying these augmentation techniques on other colonoscopy tasks such as polyp detection \& segmentation. Another interesting direction is  using the GAN generated data in semi-supervised \& unsupervised settings for colonoscopy tasks \cite{Golhar}. 

\section*{Acknowledgments}
This work was supported in part by the National Institutes of Health Trailblazer Award (No. R21 EB024700).

\printbibliography

@article{Siegel,
  title={Colorectal cancer statistics, 2020},
  author={Siegel, Rebecca L and Miller, Kimberly D and Goding Sauer, Ann and Fedewa, Stacey A and Butterly, Lynn F and Anderson, Joseph C and Cercek, Andrea and Smith, Robert A and Jemal, Ahmedin},
  journal={CA: a cancer journal for clinicians},
  volume={70},
  number={3},
  pages={145--164},
  year={2020},
  publisher={Wiley Online Library}
}

@article{Wiegering,
  title={Improved survival of patients with colon cancer detected by screening colonoscopy},
  author={Wiegering, Armin and Ackermann, Sabine and Riegel, Johannes and Dietz, Ulrich A and G{\"o}tze, Oliver and Germer, Christoph-Thomas and Klein, Ingo},
  journal={International journal of colorectal disease},
  volume={31},
  number={5},
  pages={1039--1045},
  year={2016},
  publisher={Springer}
}

@article{Nogueira,
  title={Deep neural networks approaches for detecting and classifying colorectal polyps},
  author={Nogueira-Rodriguez, Alba and Dominguez-Carbajales, Ruben and Lopez-Fernandez, Hugo and Iglesias, Agueda and Cubiella, Joaquin and Fdez-Riverola, Florentino and Reboiro-Jato, Miguel and Glez-Pena, Daniel},
  journal={Neurocomputing},
  volume={423},
  pages={721--734},
  year={2021},
  publisher={Elsevier}
}

@article{Shorten,
  title={A survey on image data augmentation for deep learning},
  author={Shorten, Connor and Khoshgoftaar, Taghi M},
  journal={Journal of big data},
  volume={6},
  number={1},
  pages={1--48},
  year={2019},
  publisher={Springer}
}

@inproceedings{Wimmer,
  title={Improving Endoscopic Decision Support Systems by Translating Between Imaging Modalities},
  author={Wimmer, Georg and Gadermayr, Michael and V{\'e}csei, Andreas and Uhl, Andreas},
  booktitle={International Workshop on Simulation and Synthesis in Medical Imaging},
  pages={131--141},
  year={2020},
  organization={Springer}
}

@inproceedings{Mathew,
  title={Augmenting colonoscopy using extended and directional CycleGAN for lossy image translation},
  author={Mathew, Shawn and Nadeem, Saad and Kumari, Sruti and Kaufman, Arie},
  booktitle={Proceedings of the IEEE/CVF Conference on Computer Vision and Pattern Recognition},
  pages={4696--4705},
  year={2020}
}

@article{Mahmood,
  title={Unsupervised reverse domain adaptation for synthetic medical images via adversarial training},
  author={Mahmood, Faisal and Chen, Richard and Durr, Nicholas J},
  journal={IEEE transactions on medical imaging},
  volume={37},
  number={12},
  pages={2572--2581},
  year={2018},
  publisher={IEEE}
}

@article{Xia,
  title={GAN inversion: A survey},
  author={Xia, Weihao and Zhang, Yulun and Yang, Yujiu and Xue, Jing-Hao and Zhou, Bolei and Yang, Ming-Hsuan},
  journal={arXiv preprint arXiv:2101.05278},
  year={2021}
}

@article{Karras_2020,
  title={Training generative adversarial networks with limited data},
  author={Karras, Tero and Aittala, Miika and Hellsten, Janne and Laine, Samuli and Lehtinen, Jaakko and Aila, Timo},
  journal={Advances in Neural Information Processing Systems},
  volume={33},
  pages={12104--12114},
  year={2020}
}

@inproceedings{Karras_2019,
  title={A style-based generator architecture for generative adversarial networks},
  author={Karras, Tero and Laine, Samuli and Aila, Timo},
  booktitle={Proceedings of the IEEE/CVF conference on computer vision and pattern recognition},
  pages={4401--4410},
  year={2019}
}

@article{omer,
  title={Designing an encoder for stylegan image manipulation},
  author={Tov, Omer and Alaluf, Yuval and Nitzan, Yotam and Patashnik, Or and Cohen-Or, Daniel},
  journal={ACM Transactions on Graphics (TOG)},
  volume={40},
  number={4},
  pages={1--14},
  year={2021},
  publisher={ACM New York, NY, USA}
}

@inproceedings{He,
  title={Deep residual learning for image recognition},
  author={He, Kaiming and Zhang, Xiangyu and Ren, Shaoqing and Sun, Jian},
  booktitle={Proceedings of the IEEE conference on computer vision and pattern recognition},
  pages={770--778},
  year={2016}
}

@article{Borgli,
  title={HyperKvasir, a comprehensive multi-class image and video dataset for gastrointestinal endoscopy},
  author={Borgli, Hanna and Thambawita, Vajira and Smedsrud, Pia H and Hicks, Steven and Jha, Debesh and Eskeland, Sigrun L and Randel, Kristin Ranheim and Pogorelov, Konstantin and Lux, Mathias and Nguyen, Duc Tien Dang and others},
  journal={Scientific data},
  volume={7},
  number={1},
  pages={1--14},
  year={2020},
  publisher={Nature Publishing Group}
}

@article{Mesejo,
  title={Computer-aided classification of gastrointestinal lesions in regular colonoscopy},
  author={Mesejo, Pablo and Pizarro, Daniel and Abergel, Armand and Rouquette, Olivier and Beorchia, Sylvain and Poincloux, Laurent and Bartoli, Adrien},
  journal={IEEE transactions on medical imaging},
  volume={35},
  number={9},
  pages={2051--2063},
  year={2016},
  publisher={IEEE}
}

@article{Li,
  title={Colonoscopy polyp detection and classification: Dataset creation and comparative evaluations},
  author={Li, Kaidong and Fathan, Mohammad I and Patel, Krushi and Zhang, Tianxiao and Zhong, Cuncong and Bansal, Ajay and Rastogi, Amit and Wang, Jean S and Wang, Guanghui},
  journal={Plos one},
  volume={16},
  number={8},
  pages={e0255809},
  year={2021},
  publisher={Public Library of Science San Francisco, CA USA}
}

@article{fid,
  title={Gans trained by a two time-scale update rule converge to a local nash equilibrium},
  author={Heusel, Martin and Ramsauer, Hubert and Unterthiner, Thomas and Nessler, Bernhard and Hochreiter, Sepp},
  journal={Advances in neural information processing systems},
  volume={30},
  year={2017}
}

@inproceedings{Zhu,
  title={Unpaired image-to-image translation using cycle-consistent adversarial networks},
  author={Zhu, Jun-Yan and Park, Taesung and Isola, Phillip and Efros, Alexei A},
  booktitle={Proceedings of the IEEE international conference on computer vision},
  pages={2223--2232},
  year={2017}
}

@inproceedings{Pau,
  title={Embedding propagation: Smoother manifold for few-shot classification},
  author={Rodriguez, Pau and Laradji, Issam and Drouin, Alexandre and Lacoste, Alexandre},
  booktitle={European Conference on Computer Vision},
  pages={121--138},
  year={2020},
  organization={Springer}
}

@article{Golhar,
  title={Improving colonoscopy lesion classification using semi-supervised deep learning},
  author={Golhar, Mayank and Bobrow, Taylor L and Khoshknab, Mirmilad Pourmousavi and Jit, Simran and Ngamruengphong, Saowanee and Durr, Nicholas J},
  journal={IEEE Access},
  volume={9},
  pages={631--640},
  year={2020},
  publisher={IEEE}
}

\clearpage
\appendix

\section*{Supplementary}

\begin{figure}[!ht]
\centering
\includegraphics[width=0.65\columnwidth]{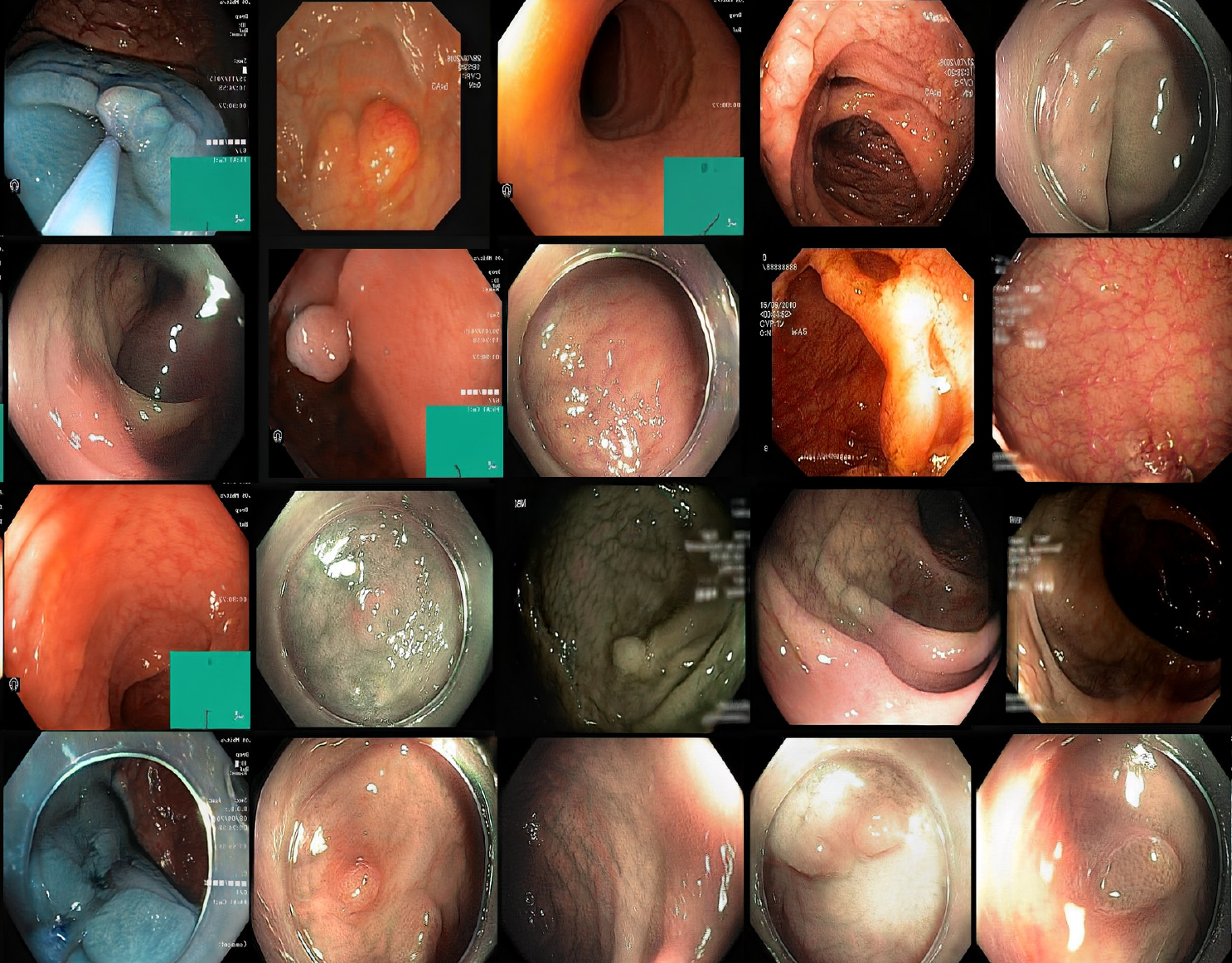}
\caption{Illustration showing realistic-looking synthetic images generated by StyleGAN-ADA with random seeds. We can observe images with variety in imaging modalities, illumination intensities, lesion shapes, vasculature patterns, etc. } \label{fig1}
\end{figure}

\begin{table}[!ht]
\caption{Breakdown of dataset used for Polyp Classification.}
\label{tab:dataset}
\centering
\resizebox{0.4\linewidth}{!}{%
\begin{tabular}{|l|l|l|l|}
\hline
\textbf{\begin{tabular}[c]{@{}l@{}}Modes $\rightarrow$ \\ Class $\downarrow$ \end{tabular}} & \textbf{WLI} & \textbf{NBI} & \textbf{Dyed} \\ \hline
\textbf{Non-neoplastic} & 179 & 152 & 0 \\ \hline
\textbf{Neoplastic }& 949 & 692 & 47 \\ \hline
\end{tabular}
}%
\end{table}

\begin{table}[!ht]
\caption{List of image classes of HyperKvasir dataset used for training StyleGAN-ADA. The last three columns indicate the source data folders for the image class. For annotating unlabeled images, if both the pre-trained classifiers (provided in the HyperKvasir paper) label the image with the same class, we assign the image the classifiers' output label.}
\centering
\resizebox{0.8\columnwidth}{!}{%
\begin{tabular}{|l|l|l|c|c|c|}
\hline
\multicolumn{1}{|c|}{\textbf{\begin{tabular}[c]{@{}c@{}}Anatomical \\ Position\end{tabular}}} & \multicolumn{1}{c|}{\textbf{Type}} & \multicolumn{1}{c|}{\textbf{\begin{tabular}[c]{@{}c@{}}Image\\ Class\end{tabular}}} & \textbf{\begin{tabular}[c]{@{}l@{}}Labeled\\ Images\end{tabular}} & \textbf{\begin{tabular}[c]{@{}l@{}}Labeled\\ Videos\end{tabular}} & \textbf{\begin{tabular}[c]{@{}l@{}}Unlabeled\\ Images\end{tabular}} \\ \hline
\multirow{14}{*}{Lower GI} & \multirow{9}{*}{Pathological Findings} & Polyps & \cmark & \cmark & \cmark \\ \cline{3-6}  
 &  & Ulcerative Colitis Grade 0-1  & \cmark & \xmark &  \xmark\\ \cline{3-6} 
 &  & Ulcerative Colitis Grade 1  & \cmark & \xmark & \xmark\\ \cline{3-6} 
 &  & Ulcerative Colitis Grade 1-2  & \cmark & \xmark & \xmark\\ \cline{3-6} 
 &  & Ulcerative Colitis Grade 2  & \cmark & \xmark & \xmark\\ \cline{3-6} 
 &  & Ulcerative Colitis Grade 2-3  & \cmark & \xmark & \xmark\\ \cline{3-6} 
 &  & Ulcerative Colitis Grade 3  & \cmark & \xmark & \xmark\\ \cline{3-6} 
 &  & Colitis  & \xmark & \cmark & \xmark \\ \cline{3-6} 
 &  & Colorectal Cancer  & \xmark & \cmark & \xmark\\ \cline{2-6} 
 & \multirow{2}{*}{Therapeutic interventions} & Dyed Lifted Polyps  & \cmark & \cmark & \cmark \\ \cline{3-6}
 &  & Dyed Resection Margins  & \xmark & \cmark & \cmark\\ \cline{2-6} 
 & \multirow{3}{*}{Quality of Mucosal Views} & BBPS-0-1  & \cmark & \cmark & \cmark\\ \cline{3-6} 
 &  & BBPS-2-3  & \cmark & \cmark & \cmark \\ \cline{3-6} 
 &  & Impacted Stool  & \cmark & \xmark & \xmark \\ \hline
\end{tabular}
}%
\end{table}


\begin{table}[!ht]
\caption{Hyperparameters used for polyp classification experiments using WLI-NBI domain transfer based data augmentation (Table 1 experiments in main paper)}
\label{tab:domain-transfer}
\centering
\resizebox{\columnwidth}{!}{%
\begin{tabular}{llllllllllllll}
\hline
\multicolumn{1}{|l}{\textbf{Type}} & \textbf{Train Set} & \textbf{\begin{tabular}[c]{@{}l@{}}Test \\ Set\end{tabular}} & \textbf{\begin{tabular}[c]{@{}l@{}}Test\\ Fold\end{tabular}} & \textbf{\begin{tabular}[c]{@{}l@{}}Learning\\ Rate\end{tabular}} & \textbf{\begin{tabular}[c]{@{}l@{}}L2\\ Penalty\end{tabular}} & \multicolumn{1}{l|}{\textbf{\begin{tabular}[c]{@{}l@{}}Dropout \\ Probability\end{tabular}}} & \textbf{Type} & \textbf{Train Set} & \textbf{\begin{tabular}[c]{@{}l@{}}Test \\ Set\end{tabular}} & \textbf{\begin{tabular}[c]{@{}l@{}}Test\\ Fold\end{tabular}} & \textbf{\begin{tabular}[c]{@{}l@{}}Learning\\ Rate\end{tabular}} & \textbf{\begin{tabular}[c]{@{}l@{}}L2\\ Penalty\end{tabular}} & \multicolumn{1}{l|}{\textbf{\begin{tabular}[c]{@{}l@{}}Dropout \\ Probability\end{tabular}}} \\ \hline
\multicolumn{1}{|l}{\multirow{5}{*}{Baseline}} & \multirow{5}{*}{NBI} & \multirow{5}{*}{NBI} & 0 & 1e-4 & 1e-3 & \multicolumn{1}{l|}{0} & \multirow{5}{*}{CycleGAN} & \multirow{5}{*}{NBI+$NBI_{transl}$} & \multirow{5}{*}{NBI} & 0 & 1e-4 & 1e-3 & \multicolumn{1}{l|}{0} \\
\multicolumn{1}{|l}{} &  &  & 1 & 1e-4 & 1e-3 & \multicolumn{1}{l|}{0} &  &  &  & 1 & 1e-4 & 1e-3 & \multicolumn{1}{l|}{0} \\
\multicolumn{1}{|l}{} &  &  & 2 & 1e-4 & 1e-2 & \multicolumn{1}{l|}{0.5} &  &  &  & 2 & 1e-4 & 1e-3 & \multicolumn{1}{l|}{0} \\
\multicolumn{1}{|l}{} &  &  & 3 & 1e-4 & 1e-3 & \multicolumn{1}{l|}{0} &  &  &  & 3 & 1e-4 & 1e-3 & \multicolumn{1}{l|}{0} \\
\multicolumn{1}{|l}{} &  &  & 4 & 1e-4 & 1e-2 & \multicolumn{1}{l|}{0.5} &  &  &  & 4 & 1e-4 & 1e-3 & \multicolumn{1}{l|}{0} \\ \hline
\multicolumn{1}{|l}{\multirow{5}{*}{Baseline}} & \multirow{5}{*}{NBI + WLI} & \multirow{5}{*}{NBI} & 0 & 1e-4 & 1e-3 & \multicolumn{1}{l|}{0} & \multirow{5}{*}{Ours} & \multirow{5}{*}{NBI + $NBI_{transl}$} & \multirow{5}{*}{NBI} & 0 & 1e-5 & 1e-2 & \multicolumn{1}{l|}{0.5} \\
\multicolumn{1}{|l}{} &  &  & 1 & 1e-4 & 1e-2 & \multicolumn{1}{l|}{0.5} &  &  &  & 1 & 1e-4 & 1e-3 & \multicolumn{1}{l|}{0} \\
\multicolumn{1}{|l}{} &  &  & 2 & 1e-4 & 1e-2 & \multicolumn{1}{l|}{0.5} &  &  &  & 2 & 1e-5 & 1e-3 & \multicolumn{1}{l|}{0} \\
\multicolumn{1}{|l}{} &  &  & 3 & 1e-4 & 1e-3 & \multicolumn{1}{l|}{0} &  &  &  & 3 & 1e-4 & 1e-3 & \multicolumn{1}{l|}{0} \\
\multicolumn{1}{|l}{} &  &  & 4 & 1e-4 & 1e-2 & \multicolumn{1}{l|}{0.5} &  &  &  & 4 & 1e-5 & 1e-3 & \multicolumn{1}{l|}{0} \\ \hline
\multicolumn{1}{|l}{\multirow{5}{*}{Baseline}} & \multirow{5}{*}{NBI+WLI} & \multirow{5}{*}{\begin{tabular}[c]{@{}l@{}}NBI\\ +WLI\end{tabular}} & 0 & 1e-4 & 1e-2 & \multicolumn{1}{l|}{0.5} & \multirow{5}{*}{CycleGAN} & \multirow{5}{*}{\begin{tabular}[c]{@{}l@{}}NBI+WLI\\ +$NBI_{transl}$+$WLI_{transl}$\end{tabular}} & \multirow{5}{*}{\begin{tabular}[c]{@{}l@{}}NBI\\ +WLI\end{tabular}} & 0 & 1e-5 & 1e-3 & \multicolumn{1}{l|}{0} \\
\multicolumn{1}{|l}{} &  &  & 1 & 1e-4 & 1e-3 & \multicolumn{1}{l|}{0} &  &  &  & 1 & 1e-4 & 1e-3 & \multicolumn{1}{l|}{0} \\
\multicolumn{1}{|l}{} &  &  & 2 & 1e-4 & 1e-3 & \multicolumn{1}{l|}{0} &  &  &  & 2 & 1e-4 & 1e-3 & \multicolumn{1}{l|}{0} \\
\multicolumn{1}{|l}{} &  &  & 3 & 1e-4 & 1e-3 & \multicolumn{1}{l|}{0} &  &  &  & 3 & 1e-4 & 1e-2 & \multicolumn{1}{l|}{0.5} \\
\multicolumn{1}{|l}{} &  &  & 4 & 1e-4 & 1e-3 & \multicolumn{1}{l|}{0} &  &  &  & 4 & 1e-4 & 1e-2 & \multicolumn{1}{l|}{0.5} \\ \hline
\multicolumn{1}{|l}{\multirow{5}{*}{Ours}} & \multirow{5}{*}{\begin{tabular}[c]{@{}l@{}}NBI+WLI\\ +$NBI_{transl}$+$WLI_{transl}$\end{tabular}} & \multirow{5}{*}{\begin{tabular}[c]{@{}l@{}}NBI\\ +WLI\end{tabular}} & 0 & 1e-5 & 1e-2 & \multicolumn{1}{l|}{0} &  &  &  &  &  &  &  \\
\multicolumn{1}{|l}{} &  &  & 1 & 1e-4 & 1e-3 & \multicolumn{1}{l|}{0} &  &  &  &  &  &  &  \\
\multicolumn{1}{|l}{} &  &  & 2 & 1e-4 & 1e-2 & \multicolumn{1}{l|}{0} &  &  &  &  &  &  &  \\
\multicolumn{1}{|l}{} &  &  & 3 & 1e-4 & 1e-2 & \multicolumn{1}{l|}{0} &  &  &  &  &  &  &  \\
\multicolumn{1}{|l}{} &  &  & 4 & 1e-5 & 1e-2 & \multicolumn{1}{l|}{0.5} &  &  &  &  &  &  &  \\ \cline{1-7}
 &  &  &  &  &  &  &  &  &  &  &  &  &  \\
 &  &  &  &  &  &  &  &  &  &  &  &  &  \\
 &  &  &  &  &  &  &  &  &  &  &  &  &  \\
 &  &  &  &  &  &  &  &  &  &  &  &  &  \\
 &  &  &  &  &  &  &  &  &  &  &  &  & 
\end{tabular}
}%
\end{table}

\begin{table}[!ht]
\caption{Hyperparameters for polyp classification experiments using interpolation based data augmentation (Table 2 experiments in main paper)}
\centering
\resizebox{\columnwidth}{!}{%
\begin{tabular}{|lllllll|lllllll|}
\hline
\textbf{Type} & \textbf{\begin{tabular}[c]{@{}l@{}}Train\\ Set\end{tabular}} & \textbf{\begin{tabular}[c]{@{}l@{}}Test\\ Set\end{tabular}} & \textbf{\begin{tabular}[c]{@{}l@{}}Test\\ Fold\end{tabular}} & \textbf{\begin{tabular}[c]{@{}l@{}}Learning\\ Rate\end{tabular}} & \textbf{\begin{tabular}[c]{@{}l@{}}L2 \\ penalty\end{tabular}} & \textbf{\begin{tabular}[c]{@{}l@{}}Dropout\\ Probability\end{tabular}} & \textbf{Type} & \textbf{\begin{tabular}[c]{@{}l@{}}Train\\ Set\end{tabular}} & \textbf{\begin{tabular}[c]{@{}l@{}}Test\\ Set\end{tabular}} & \textbf{\begin{tabular}[c]{@{}l@{}}Test\\ Fold\end{tabular}} & \textbf{\begin{tabular}[c]{@{}l@{}}Learning\\ Rate\end{tabular}} & \textbf{\begin{tabular}[c]{@{}l@{}}L2 \\ penalty\end{tabular}} & \textbf{\begin{tabular}[c]{@{}l@{}}Dropout\\ Probability\end{tabular}} \\ \hline
\multirow{5}{*}{Baseline} & \multirow{5}{*}{WLI} & \multirow{5}{*}{WLI} & 0 & 1e-4 & 1e-2 & 0.5 & \multirow{5}{*}{Baseline} & \multirow{5}{*}{NBI} & \multirow{5}{*}{NBI} & 0 & 1e-4 & 1e-3 & 0 \\
 &  &  & 1 & 1e-4 & 1e-3 & 0 &  &  &  & 1 & 1e-4 & 1e-3 & 0 \\
 &  &  & 2 & 1e-4 & 1e-3 & 0 &  &  &  & 2 & 1e-4 & 1e-2 & 0.5 \\
 &  &  & 3 & 1e-4 & 1e-2 & 0.5 &  &  &  & 3 & 1e-4 & 1e-3 & 0 \\
 &  &  & 4 & 1e-4 & 1e-3 & 0 &  &  &  & 4 & 1e-4 & 1e-2 & 0.5 \\ \hline
\multirow{5}{*}{Ours} & \multirow{5}{*}{\begin{tabular}[c]{@{}l@{}}WLI+\\ $WLI_{interp}$\end{tabular}} & \multirow{5}{*}{WLI} & 0 & 1e-4 & 1e-2 & 0.5 & \multirow{5}{*}{Ours} & \multirow{5}{*}{\begin{tabular}[c]{@{}l@{}}NBI+\\ $NBI_{interp}$\end{tabular}} & \multirow{5}{*}{NBI} & 0 & 1e-5 & 1e-2 & 0 \\
 &  &  & 1 & 1e-5 & 1e-2 & 0.5 &  &  &  & 1 & 1e-4 & 1e-2 & 0.5 \\
 &  &  & 2 & 1e-5 & 1e-2 & 0.5 &  &  &  & 2 & 1e-5 & 1e-4 & 0 \\
 &  &  & 3 & 1e-4 & 1e-4 & 0 &  &  &  & 3 & 1e-4 & 1e-3 & 0 \\
 &  &  & 4 & 1e-4 & 1e-3 & 0 &  &  &  & 4 & 1e-4 & 1e-3 & 0 \\ \hline
\end{tabular}
}%
\end{table}
\end{document}